\documentstyle[pra,aps,psfig,twocolumn]{revtex}

\include{pdftex}

\begin{document}

\draft % needed to display pacs
%
% USEFUL DEFINITIONS:
%
\def\sn{{\rm sn}} \def\cn{{\rm cn}} \def\dn{{\rm dn}}
\def\sech{{\rm sech}}
\def\ds{\displaystyle}
\def\FigSize{8.40cm}
\def\root{}
\def\p{\partial}
\def\pp#1#2{\ds{\ds\p #1 \over \ds\p #2}}
\def\mybf{}
%
%%%%%%%%%%%%%%%%%%%%%%%%%%%%%%%%%%%%%%%%%%%%%%%%%%%%%%%%%%%%%%%%%%%%%%%%
%% FIGURE MACRO: \oneFIG{name}{width[cm]}{caption} %%%%%%%%%%%%%%%%%%%%%
%%%%%%%%%%%%%%%%%%%%%%%%%%%%%%%%%%%%%%%%%%%%%%%%%%%%%%%%%%%%%%%%%%%%%%%%
\def\oneFIG#1#2#3{
\begin{figure}[ht]
 \vskip -0.2cm
 \centerline{\psfig{figure=\root#1,width=#2,silent=}}
 \vskip 0.1cm 
 \caption[]{#3\label{#1}} 
\end{figure} 
\vskip -0.2cm}
%%%%%%%%%%%%%%%%%%%%%%%%%%%%%%%%%%%%%%%%%%%%%%%%%%%%%%%%%%%%%%%%%%%%%%%%
\def\twoFIG#1#2#3#4#5{
\begin{figure}[ht]
 \vskip -0.2cm
 \centerline{\psfig{figure=\root#1,width=#3,silent=}}
 \vskip #4
 \centerline{\psfig{figure=\root#2,width=#3,silent=}}
 \vskip 0.1cm 
 \caption[]{#5\label{#1}} 
\end{figure} 
\vskip -0.2cm}
%%%%%%%%%%%%%%%%%%%%%%%%%%%%%%%%%%%%%%%%%%%%%%%%%%%%%%%%%%%%%%%%%%%%%%%%

\wideabs{

\null\vskip 0.51cm

\title{\bf Localized breathing oscillations of Bose-Einstein
           condensates in periodic traps.}

\author{
R.~Carretero-Gonz\'alez$^{1,}$\cite{rcg:email} and
and K.~Promislow$^2$}

\address{
$^1$Nonlinear Dynamical Systems Group\cite{NLDS:web},
Department of Mathematics \& Statistics,\\ %\cite{SDSU:web},\\
San Diego State University, San Diego, CA 92182-7720, USA.\\
$^2$Department of Mathematics \& Statistics\cite{SFU:web},
Simon Fraser University, Burnaby, BC, V5A 1S6, Canada\\
}

\date{{\em Phys.\ Rev.\ A}, {\bf 66}, 033610 (2002))\\
%      \today.
      (For related work please visit:
      http://www.rohan.sdsu.edu/$\sim$rcarrete/publications/}

\maketitle

%\vskip -3cm
%Accepted for publication in {\em Physical Review A}, 2002.
%\vskip 3cm

%-------------------------------------------------------------------------
\begin{abstract} 
%       10        20        30        40        50        60        70
%
We demonstrate the existence of localized oscillatory breathers for
quasi-one-dimensional Bose-Einstein condensates confined in periodic 
potentials. The breathing behavior corresponds to position-oscillations
of individual condensates about the minima of the potential lattice.
%Localized oscillations are identified with homoclinic tangles of a
%reduced two-dimensional map on the oscillation amplitudes. 
We deduce the structural stability of the localized oscillations 
from the construction.  The stability is confirmed numerically
for perturbations to the initial state of the condensate, to
the potential trap, as well as for external noise.
We also construct periodic and chaotic extended
oscillations for the chain of condensates. 
All our findings are verified by direct numerical integration of the
Gross-Pitaevskii equation in one dimension.
%The regimes we consider are compatible
%with current physical parameter values, opening the possibility for
%experimental corroboration.
%
\end{abstract}
%-------------------------------------------------------------------------

%03. Quantum mechanics, field theories, and special relativity 
%  03.75.-b Matter waves
%  03.75.Fi Phase coherent atomic ensembles; quantum condensation phenomena
%
%42. Optics
%  42.65.-k Nonlinear optics
%
%52. Physics of plasmas and electric discharges
%  52.35.Mw Nonlinear phenomena: waves, wave propag., and other interactions
%
%05. Statistical physics, thermodynamics, and nonlinear dynamical systems
%  05.45.-a Nonlinear dynamics and nonlinear dynamical systems 
%
%63. Lattice dynamics 
%  63.20.Pw Localized modes
%
}%end wideabs

%\tableofcontents

New techniques for generating Bose-Einstein condensates (BECs) have opened 
the door to the investigation of a wide range of phenomena and development
of concrete applications such as atomic interferometers and atom lasers.
Recent focus has been on BECs trapped in periodic magnetic/optical
traps \cite{Anderson-Kasevich-Jaksch,Berg:98,Trombettoni:01}, the
so-called optical lattice \cite{Jessen-Petsas}.
The addition of a periodic potential to the 
BEC opens the possibility to study exciting dynamical phenomena 
of BECs grown in optical traps, such as
Bloch oscillations and the Josephson effect
\cite{Anderson-Kasevich-Jaksch,Berg:98,Trombettoni:01}.
One approach for the study of chains of weakly coupled condensates, in
the tight binding approximation, reduces the dynamics of the BECs to
the discrete Nonlinear Schr\"odinger equation (DNLS) 
\cite{Berg:98,Trombettoni:01,Javanainen:99,Chiofalo:00,Konotop:01}.
Within this approach, each BEC wave function, centered at the
minima of the potential, is modeled by the {\em same} {\em stationary}
wave function with a time dependent amplitude.
This approximation leads to the DNLS for the local
time dependent amplitudes. The dynamics described by
this approach allows for variations in the local
population of atoms, but are limited to a fixed local
distribution of the condensed atoms.
In this manuscript we use a more general approach where the
local wave function evolves within an ansatz family,
allowing for a wider range of dynamical behavior. 
Using the treatment presented here it is possible to include
variations in the local populations as well as in the {\em shape}
of the local density: height, width, position, phase, velocity, etc.
In particular, we focus our attention on position-oscillations of 
each {\em local} BEC density about its potential minimum by considering 
a constant population at each trough of the periodic potential.
We model the coupled chain of oscillating condensates by
a nonlinear lattice (a Toda lattice with on-site effective potentials)
for the oscillation amplitudes. We establish the existence of
global and localized oscillations of the coupled condensates,
and demonstrate the ability
of condensates to store macroscopic information in localized
regions of the periodic trap.

The wave function of a dilute BECs at low temperature
is governed by the Gross-Pitaevskii equation (GPE) 
\cite{Gross-Pitaevskii} 
%\cite{Gross:61,Pitaevskii:61}. 
%
\begin{equation}\label{GPE}
{\rm i} \hbar \pp{\psi}{t} = \left(-{\hbar^2 \nabla^2 \over 2m} +
V_{\rm ext}(\vec{r})+ g_0 |\psi|^2\right)\psi,
\end{equation}
where $\psi=\psi(\vec{r},t)$ is the condensate wave function normalized
with respect to the total number of condensed atoms $\cal{N}$. The
atom-atom interactions (considered only binary due to the low temperature
assumption) are accounted for by the coupling constant \cite{Dalfovo:99}
%
%\begin{equation}\label{g}
$$
g_0={4 \pi \hbar^2 a \over m}
$$
%\end{equation}
%
where $a$ is the $s$-wave scattering length and 
$m$ the atomic mass. The external potential $V_{\rm ext}(\vec{r})$ 
is given by the sum of the confining potential $V_{\rm conf}(\vec{r})$ 
and a periodic optical potential $V_{\rm per}(\vec{r})$:
$$
V_{\rm ext}(\vec{r})=V_{\rm conf}(\vec{r})+V_{\rm per}(\vec{r}).
$$
%
%%%%%%%%%%%%%%%%%%%%%%%%%%%%%%%%%%%%%%%%%%%%%%%%%%%%%%%%%%%%%%%%%%%%%%%%
\oneFIG{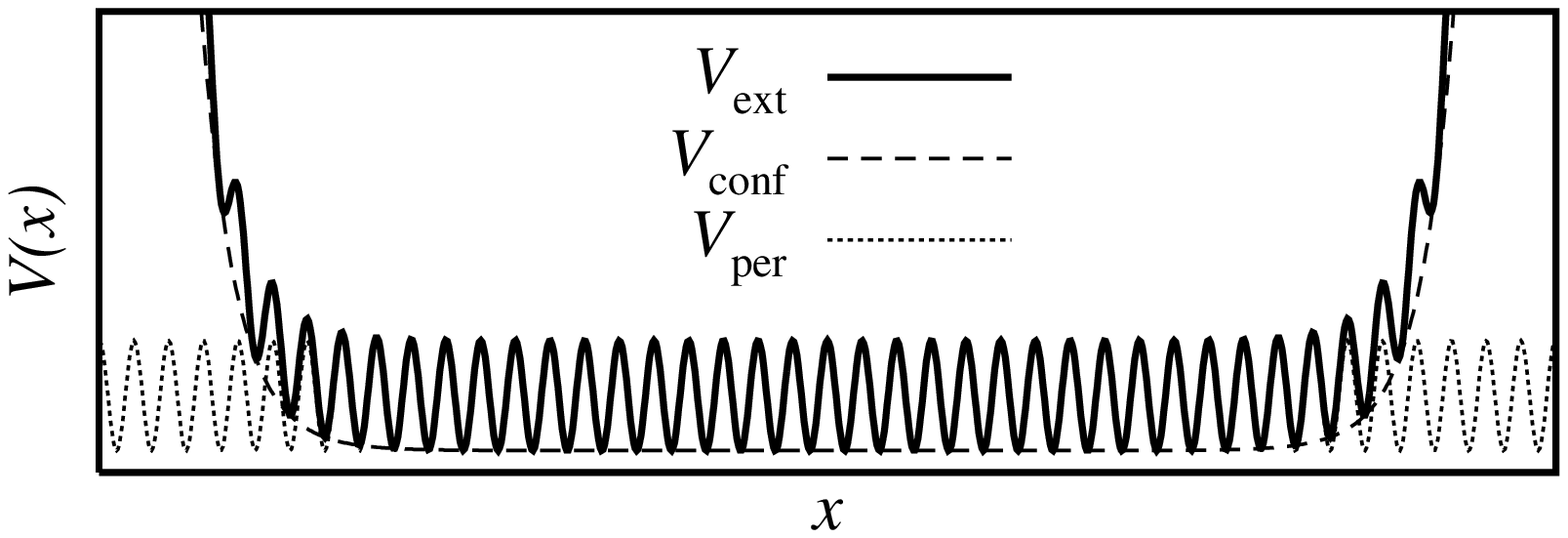}{\FigSize}{External potential $V_{\rm ext}$ with 
confining ($V_{\rm conf}$) and periodic ($V_{\rm per}(x)$) properties.}
%%%%%%%%%%%%%%%%%%%%%%%%%%%%%%%%%%%%%%%%%%%%%%%%%%%%%%%%%%%%%%%%%%%%%%%%
%
\noindent We consider confining traps that give rise to 
the so-called cigar-shaped condensates whose
longitudinal direction is much larger than
their transverse dimension, which is of the order of the healing length.
In this cigar-shaped limit it is possible to reduce
the GPE equation (\ref{GPE}) to its one-dimensional analogue
\cite{Jackson-PerezGarcia-Key-Dekker}.
%It is also possible to reduce the
%the GPE to its one-dimensional analogue for a stack
%of `flat pancakes' of condensates formed from two superimposed
%optical lattices \cite{Konotop:pc}.
Our study will be limited to the one-dimensional model of attractive BECs.
Extension of our results to the repulsive BEC in one and higher 
dimensions will be reported elsewhere.
Our model applies to the dynamics of a chain of 
quasi-identical condensates, in an infinite confining trap.
%a flat center portion fitting several tens of the periodic potential
%period (see fig \ref{fig1.ps}).
%
In practice, the harmonic external trap has a flat central portion supporting
several hundreds of periods of the periodic potential, which is sufficient
to satisfy the desired near-periodic potential (see fig \ref{fig1.ps}). 
It is also possible to load the condensate onto $V_{\rm ext}$ and then
adiabatically remove  $V_{\rm conf}$, leaving only the periodic potential.

After rescaling, the BEC trapped inside a periodic
potential $V$ is governed by the nonlinear Schr\"odinger equation (NLSE):
\begin{equation}\label{NLSE}
{\rm i} u_t + {1 \over 2} u_{xx} \pm |u|^2 u = V(x) u.
\end{equation}
Here $|u(x,t)|^2$ corresponds to the rescaled density of the
condensate, $V(x)$ is the rescaled potential and the $\pm$ sign 
corresponds to attractive ($+$) and repulsive ($-$) atoms. 
%The adimensionalization of (\ref{NLSE}) is achieved by
%$t\rightarrow (4\pi\hbar|a|{\cal{N}}/m\Omega)\,t$,
%$x\rightarrow (\Omega/4\pi|a|{\cal{N}})^{1/2}\,x$,
%$u\rightarrow (\hbar\Omega)^{-1/2}\,u$ and
%$V\rightarrow (m\Omega/4\pi\hbar|a|{\cal{N}})\,V$, where ${\cal{N}}$ is
%the number of atoms in a the volume $\Omega$, $m$ is the mass
%of a single atom and $a$ is the $s$-wave scattering length
%\cite{Bernard:01}. 
%
%The nondimensionalization is obtained by scaling
%$t\rightarrow \Omega\,t$,
%$x\rightarrow (\hbar/2 \Omega)^{-1/2}\,x$ and
%$|u|^2\rightarrow (4\pi\hbar {\cal{N}} |a|/m\Omega)\,|u|^2$ where ${\cal{N}}$ 
%is the number of condensed atoms, $m$ is the atomic mass,
%$\Omega$ is the frequency of the periodic trap 
%and $a$ is the $s$-wave scattering length \cite{Ruprecht:95}. 
%
%t -> {t / \Omega}
%x -> x (\hbar / m \Omega)^{1/2}\,x
%|u|^2 -> (m \Omega / 4 \pi \hbar |a|) \, |u|^2
%Remove $\cal{N}$!!!
%
The nondimensionalization is obtained by scaling
$t\rightarrow {t / \Omega}$,
$x\rightarrow (\hbar / m \Omega)^{1/2}\,x$ and
$|u|^2\rightarrow (m \Omega / 4 \pi \hbar |a|)\,|u|^2$ where $m$ is the 
atomic mass, $\Omega$ is the frequency of the periodic trap 
and $a$ is the $s$-wave scattering length \cite{Ruprecht:95}. 
We consider a periodic potential $V(x)$, with well spaced troughs,
constructed by `concatenating' single well potentials
$V_j(x)=V_0\,\tanh^2(x-\xi_{0;j})$:
%
%\begin{equation}\label{V(x)}
$$
V(x) = \sum_{j=-N}^{N}{V_j(x)}-2N\,V_0.
%V(x) = \sum{V_j(x)}.
$$
%\end{equation}
%
The potentials are located at $\xi_{0;j} = j\,R$ where
$R$ is the potential well spacing. The specific shape of the 
potentials $V_j$ is not central to the construction.
However, our approximations require that neighboring condensates,
grown in adjacent wells, have small overlap. This regime corresponds 
to large enough $R$ or to a small number of atoms per potential well.
The large $R$ limit can be achieved by tuning the wavenumber of the 
optical trap. Alternatively, new advances in the
manipulation of condensates permit the construction of traps with 
arbitrary spatial patterns by imprinting the pattern in a crystal 
by a grating method or by lithography \cite{Hansch-group}.

The present manuscript addresses the dynamics of a chain
of coupled attractive condensates,
ie.\ {\em bright} solitons. For repulsive atoms ---{\em dark} solitons
\cite{Busch-Burger}--- a similar approach is possible and the
detailed analysis will be carried elsewhere. Current typical 
lifetimes for attractive condensates range from hundreds to
tenths of a second before collapse.
These lifetimes correspond, for typical experimental
parameters in $^{85}$Rb \cite{Bernard:01}, to $10^3$ rescaled time
units, long enough to observe 
10--20 oscillation periods. Moreover, new 
microtrap techniques \cite{Hansch-group} and hollow lasers traps
\cite{Schiffer-Wright} hold out the possibility to increase considerably
the lifetime of attractive condensates \cite{RCG:note2},
permitting time to prepare the train of condensates with the desired
phase configuration.
It is also important to note that collapse for attractive
condensates can be arrested by controlling the number of atoms.
Indeed, if the number of atoms in {\em each} potential trough is 
smaller than the critical number $N_{\rm cr}$ associated
with the local potential, collapse is avoided 
%($N_{\rm cr}$ is of the order of 1000 for $^{7}$Li) 
\cite{Bradley:97,Dalfovo:99}. 

The steady state for a single condensate inside a single well 
potential $V_j$ is a NLSE soliton solution of the form
\cite{rcg:BEC-attractive}
\begin{equation}\label{1solBEC}
u_{0;j}(x,t)= \sqrt{1-V_0}\,\, \sech(x-\xi_{0;j}) \, e^{{\rm i} (1/2-V_0) t}.
\end{equation}
We consider the dynamics of a train of condensates
%u(x,t) =\sum_{j=-\infty}^{\infty}{u_{0;j}(x,t)}.
$u(x,t) =\sum{u_{0;j}(x,t)}$ centered at the potential troughs
$\xi_{0;j}$.
For simplicity of presentation, we approximate the overall dynamics, 
focusing on two major contributions: a) internal dynamics due to interactions 
between $u_{0;j}$ and $V_j$ and b) tail-tail interactions between 
consecutive solitons. We address each of these
terms individually and use the linear superposition afforded by
soliton perturbation theory \cite{Karpman:81a}. 
Within our approximations (small tail-tail overlap) 
it is sensible to discard interactions between $u_{0;j}$ and
$V_k$ for $j\not=k$ and interactions between non-nearest neighbor
condensates \cite{Karpman:81a}.

The evolution, for small perturbations, of the steady state soliton 
$u_{0;j}$ inside its on-site potential $V_j$ can be approximated by an
oscillating soliton ansatz
%
%\begin{equation}\label{ansatz}
$
u_{j}(x,t)=u_{0;j}(x-\xi_j,t) 
$
%\end{equation}
%
with constant height and width \cite{Scharf-Bishop-Perez-Garcia}. 
The position $\xi_j(t)$ for each condensate is then described 
by a particle inside an effective potential:
\begin{equation}\label{xipp}
\ddot\xi_j = -V'_{\rm eff}(\xi_j-\xi_{0;j}).
\end{equation}
{\mybf
The effective potential $V_{\rm eff}$ may be obtained by soliton 
perturbation techniques or alternatively by demanding
that the evolution of $\xi_j(t)$ in (\ref{1solBEC}) respects the
invariance of the the NLSE (\ref{NLSE}) energy
\begin{equation}\label{E}
E=\ds\int_{-\infty}^{+\infty}\left[
\frac{1}{2} |u_x|^2 -\frac{1}{2}|u|^4 +{|u|^2 V(x)}\right] dx.
\end{equation}
%
%is a constant of motion for the NLSE (\ref{NLSE}), namely
%$dE/dt=0$. By using the ansatz (\ref{1solBEC}) together with the 
%conservation of energy, it is straightforward to prove that
%the effective potential defined through (\ref{xipp}) satisfies
It is well known \cite{Scharf-Bishop-Perez-Garcia} that the effective 
potential
\begin{equation}\label{overlapping}
V_{\rm eff}(\xi) \propto 
\int_{-\infty}^{+\infty}{|u_{0,j}(x-\xi)|^2 \,\,V_j(x) dx},
\end{equation}
is proportional to the overlapping integral between the displaced 
BEC density (\ref{1solBEC}) and the on-site potential.
Note that in (\ref{overlapping}) the on-site potential
$V_j$ is centered at $x=\xi_{0;j}$ and the
BEC density $|u_{0;j}(x-\xi)|^2$ is centered at 
$x=\xi_{0;j}+\xi$ (see (\ref{1solBEC})).
The exact form of $V_{\rm eff}$ depends upon the on-site potential. 
For the particular case under consideration, 
$V_j(x)=V_0\,\tanh^2(x-\xi_{0;j})$, the
effective potential admits the expansion
%$V_{\rm eff}(x)/V_0 = {4\over 15}x^2 - {4\over 63}x^4 + o(x^6)$.
$V_{\rm eff}(x)\propto {4\over 15}x^2 - {4\over 63}x^4 + o(x^6)$.
More generally, the effective potential can be approximated, for
small oscillations, by
\begin{equation}\label{Veff}
V_{\rm eff}(x) ={\alpha\over 2}\,x^2+{\beta\over 4}\,x^4 + o(x^6),
%V'_{\rm eff}(x) ={\alpha}\,x+{\beta}\,x^3 + o(x^5).
\end{equation}
where $\alpha$ and $\beta$ encode the shape information of $V_j$.
In particular, even symmetry of $V_{\rm eff}$ is inherited from $V_j$.
%
%{Delete rest of paragraph--a bit redundant}
%It is important to mention that,
%although the numerical results presented below correspond to
%on-site potentials of the form $V_j(x)=V_0\,\tanh^2(x-\xi_{0;j})$
%(ie.\ $\alpha=8/15$ and $\beta=16/63$),
%our analysis allows for more general on-site symmetric potentials 
%for which the values of $\alpha$ and $\beta$ are determined
%from the series expansion of the overlapping integral (\ref{overlapping}).
}

Tail-tail interactions for neighboring condensates result in
a complex version of the Toda lattice involving position and phases
\cite{Gerdjikov-Arnold}. 
%\cite{Gerdjikov:97,Arnold:99}. 
It is possible to further reduce the dynamics under the assumption 
that relative phases for consecutive condensates are constant.
This is a reasonable approximation in the regime where
the solitons are kept well apart \cite{RCG:note1}. 
{\mybf
In particular we consider the case of a $\pi$ phase shift
between consecutive condensates, since in the steady-state
($\xi_j=0$) this configuration reduces to the Jacobi
elliptic cosine which is known to be stable \cite{rcg:BEC-attractive}. 
In contrast, a chain of condensates with zero phase shift reduces to the 
third Jacobi elliptic function (dn), which is unstable 
\cite{rcg:BEC-attractive}.
}
With a relative phase of $\pi$ 
%---chosen for stability considerations \cite{rcg:BEC-attractive}---
the tail-tail interactions reduce
to a {\em real} Toda lattice (TL) \cite{Toda:book} on the positions:
$\ddot\xi_j = 4\,(e^{-(\xi_{j}-\xi_{j-1})}-e^{-(\xi_{j+1}-\xi_{j})})$.
In practice, the $\pi$ phase shift can be implemented by 
phase design on the initial configuration of the condensates
\cite{Reinhardt:00}.
From now on we consider that the amplitudes of oscillation of the
condensates are small. Thus, since the steady state is stable, we
eliminate the possibility of a condensate hopping to a neighboring 
lattice trough.

%%%%%%%%%%%%%%%%%%%%%%%%%%%%%%%%%%%%%%%%%%%%%%%%%%%%%%%%%%%%%%%%%%%%%%%%
\oneFIG{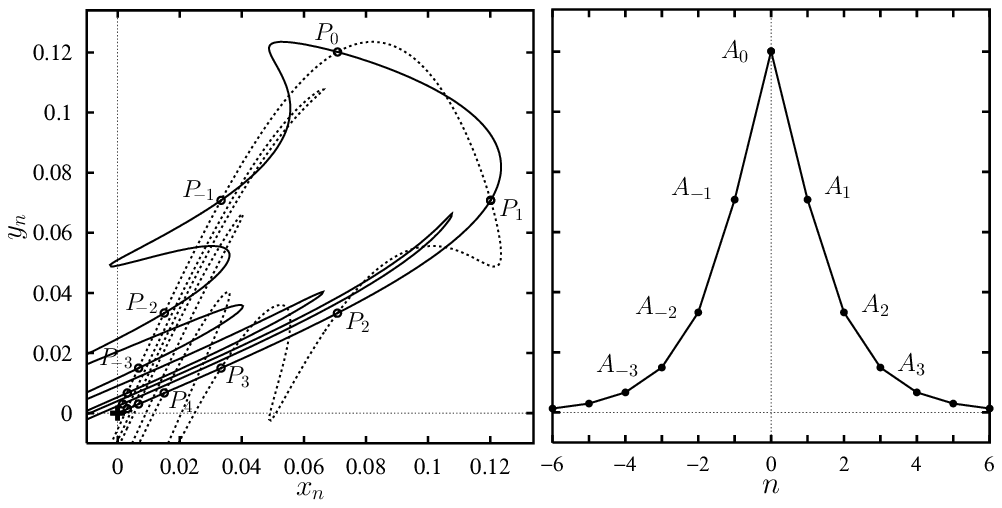}{\FigSize}{Left: Homoclinic tangle
for the two-dimensional map (\ref{2DM}).
The homoclinic orbit $\{...,P_{-1},P_{0},P_{1},...\}$ belongs to
the stable (solid line) and unstable (dash line) manifolds.
Right: The corresponding configuration for the oscillation amplitudes.
($R=10$, $\omega=17.671$ and $V_0=0.1$.)}
%%%%%%%%%%%%%%%%%%%%%%%%%%%%%%%%%%%%%%%%%%%%%%%%%%%%%%%%%%%%%%%%%%%%%%%%

Exploiting the linearity of soliton perturbation theory we combine 
the on-site potential (\ref{xipp}) with the tail-tail interactions to
find a lattice differential equation (LDE) on the condensate 
positions
\begin{equation}\label{LDE}
\ddot\xi_j = 
4 \left(e^{-(\xi_{j}-\xi_{j-1})}-e^{-(\xi_{j+1}-\xi_{j})}\right)
- V'_{\rm eff}(\xi_j-\xi_{0,j}).
\end{equation}
To find oscillatory solutions to (\ref{LDE}) we use an oscillating ansatz 
\cite{Kivshar-Flach-Bountis}, writing the position of the $j$th condensate 
as a combination of oscillatory modes
$\xi_j(t) = \sum_{k}{A_j(k)\, \cos(k\omega t)}$
centered at $A_j(0)=\xi_{0,j}$.
For small oscillations a {\em one-mode} ansatz
\begin{equation}\label{osc-ansatz}
\xi_j(t) = \xi_{0,j}+A_j\,\cos(\omega t)
\end{equation}
is sufficient to capture the essential dynamics.
We substitute (\ref{osc-ansatz}) into (\ref{LDE}),
expand and match terms to find a recurrence
relation between the oscillation amplitudes $A_j$:
\begin{equation}\label{recurrence}
A_{j+1} = (a+b\,A_{j}^2)\, A_j - A_{j-1},
\end{equation}
with $a = 2-\omega^2+ \alpha\,e^{R}/4$ and $b = 3\,\beta\,e^{R}/16$.
We introduce $y_n=A_n$ and $x_n=A_{n-1}$ and recast
the second order recurrence relation (\ref{recurrence}) as 
the 2D map $M$:
\begin{equation}\label{2DM}
%M:\,\, (x_{n+1},y_{n+1}) = (y_n,(a+b\,y_{n}^2)\, y_n - x_n).
%\begin{array}{rcl}
%(x_{n+1},y_{n+1})&=&M(x_{n},y_{n})\\[1.0ex]
%                 &=&(y_n,(a+b\,y_{n}^2)\, y_n - x_n).
%%(x_{n+1},y_{n+1})=M(x_{n},y_{n})=(y_n,(a+b\,y_{n}^2)\, y_n - x_n).
M: \left\{
\begin{array}{rcl}
x_{n+1}&=& y_n \\[1.0ex]
y_{n+1}&=& (a+b\,y_{n}^2)\, y_n - x_n
\end{array}
\right.
\end{equation}
Up to the approximations made to obtain equation (\ref{recurrence}), 
solutions of (\ref{2DM}) prescribe the amplitudes
representing oscillatory solutions for the condensates.
We remark that the oscillation frequency $\omega$ and the form 
of the on-site potential are incorporated 
through the parameters $a$ and $b$ of (\ref{recurrence}). In particular, 
one expects families of oscillatory solutions parameterized by their 
frequency $\omega$. 
%It is worth mentioning that the iteration
%index $n$ in (\ref{2DM}) corresponds to the site
%index $j$ for the amplitude of the oscillation (\ref{osc-ansatz}).

%%%%%%%%%%%%%%%%%%%%%%%%%%%%%%%%%%%%%%%%%%%%%%%%%%%%%%%%%%%%%%%%%%%%%%%%%
\oneFIG{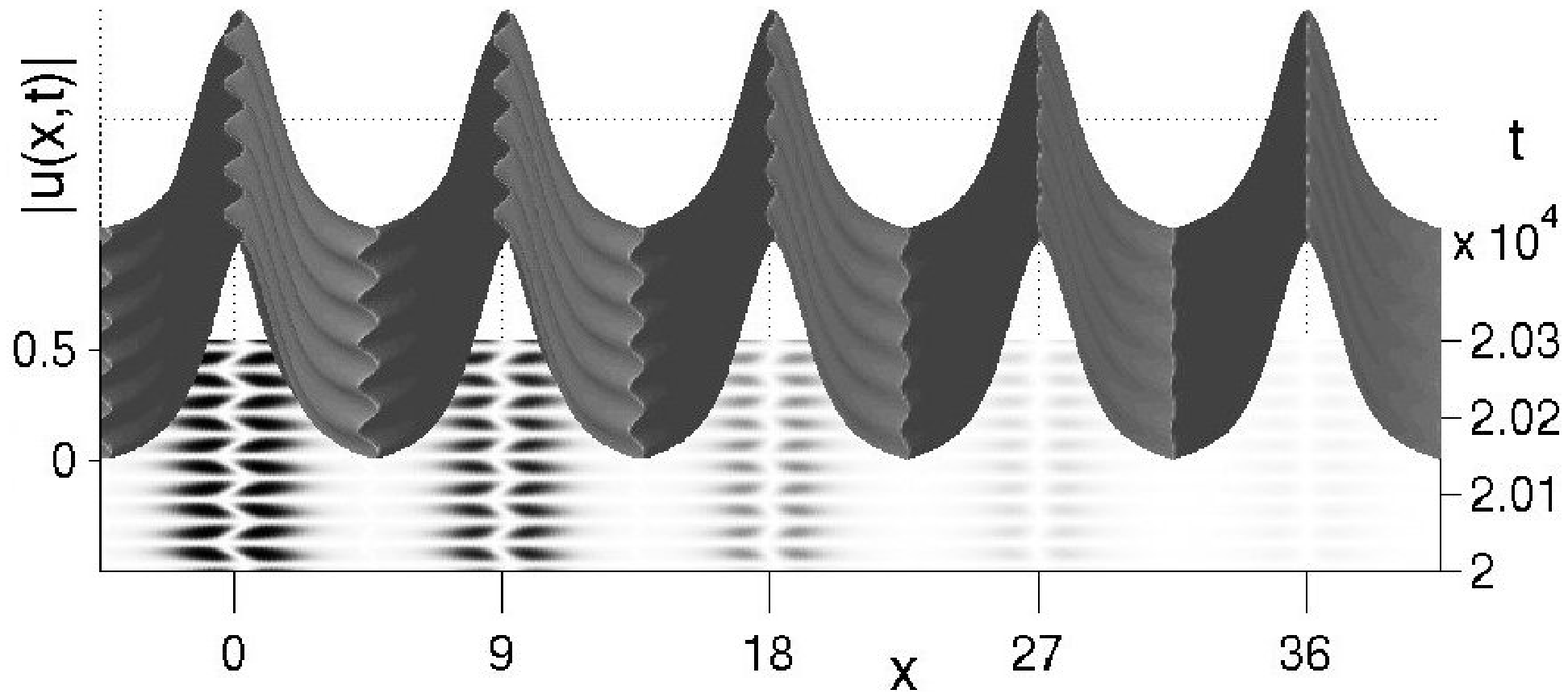}{\FigSize}{Localized breathing oscillation 
in a chain of weakly coupled condensates in a periodic potential. 
This localized oscillation is obtained by full numerical solution of 
the Gross-Pitaevskii equation with an initial condition prescribed
by our dynamical reduction.
Only condensates with index 0 through 4
are shown (the oscillations are symmetric with respect to the 
condensate with index 0). 
The bottom plane depicts $u_t$
---darker areas corresponds to regions where
$u(x,t)$ undergoes greater temporal variation.
($R=9$, $\omega\approx 0.11388$ and $V_0=0.025$.)}
%(parameter values as in Fig.\ \ref{fig2.ps}.)}
%%%%%%%%%%%%%%%%%%%%%%%%%%%%%%%%%%%%%%%%%%%%%%%%%%%%%%%%%%%%%%%%%%%%%%%%

We are interested in constructing {\em localized breathers}
(localized oscillations) for the condensate dynamics. These solutions
correspond to orbits homoclinic to the origin for the recurrence 
map (\ref{2DM}). The corresponding orbits satisfy $A_0\not=0$ and
$\lim_{n\rightarrow\pm\infty} A_n =0$ with an exponential decay rate. 
A necessary condition for the existence of a homoclinic orbit
is hyperbolicity of the origin.
This is satisfied for all $|a| >2$ or equivalently 
$\left|2-\omega^2+ \alpha\,{e^{R}/ 4}\right|>2$. This condition
provides constraints on the inter-site spacing $R$ and 
breather frequency $\omega$ for a given on-site trap $V_j$,
which determines $\alpha$. 
We find numerically, for reasonable values of $R$ ($R\approx$
a few condensate widths), that the stable ($W^s$) and unstable 
manifolds ($W^u$) of the origin intersect in a homoclinic 
tangle (see left panel in Fig.\ \ref{fig2.ps}).
{\mybf
Consider the trajectory of the two-dimensional map $M$ that starts
at the intersection point $P_0\in W^s\cap W^u$
(see left panel in Fig.\ \ref{fig2.ps}). 
From the invariance
of the stable and unstable manifolds, each forward and backward
iteration of $P_0$, labeled $\{ \ldots, P_{-2}, P_{-1}, P_0, P_1,
P_2, \ldots\}$ on  Fig.\ \ref{fig2.ps}, 
lies in the intersection. From area orientation it follows that 
this trajectory takes each second intersection \cite{RCG:note3}.
The amplitudes for the localized oscillations are then given by the
ordinates ($y_n=A_n$) of the homoclinic orbit (see right panel in Fig.\
\ref{fig2.ps}).
}

The homoclinic orbit of the 2D map induces a breather on the condensate
positions through the ansatz (\ref{osc-ansatz}).
An example of such localized oscillations for the condensates
is depicted in Fig.\ \ref{fig3.ps}, in which
a central condensate oscillates with a maximal amplitude $A_0$ and the 
condensates on either side oscillate with amplitudes $A_{\pm n}$ which 
decrease exponentially in increasing $n$ (see right panel in 
Fig.\ \ref{fig2.ps}). The asymptotic decay rate for 
the oscillations ($\lambda_-^{|n|}=\lambda_+^{-|n|}$) is prescribed 
by the eigenvalues at the origin  $\lambda_\pm = (a\pm \sqrt{a^2-4})/2$.
%($\lambda_- \simeq 0.4448$ in Fig.\ \ref{fig2.ps}).
Note that, because (\ref{2DM}) has $x\!\!\leftrightarrow\!\! y$ symmetry, the
breather is symmetric with respect to the central condensate 
($\lambda_- =\lambda_+^{-1}$).
We stress that the solution depicted in figure 
\ref{fig3.ps} is obtained by numerical integration of
the full Gross-Pitaevskii equation (\ref{GPE}) from initial
conditions prescribed by the homoclinic orbit of our reduced
2D map. Due to the approximations used in our
approach, orbits of the reduced 2D map are not {\em exact}
solutions of the full GPE equation(\ref{GPE}). Nevertheless,
if we restrict our attention to small oscillations for weakly
coupled BECs, there is a good correspondence between the reduced
dynamics and the original Gross-Pitaevskii equation. This
correspondence is reinforced by the structural stability of the
orbits for the reduced 2D map (see below).

%%%%%%%%%%%%%%%%%%%%%%%%%%%%%%%%%%%%%%%%%%%%%%%%%%%%%%%%%%%%%%%%%%%%%%%%%
\oneFIG{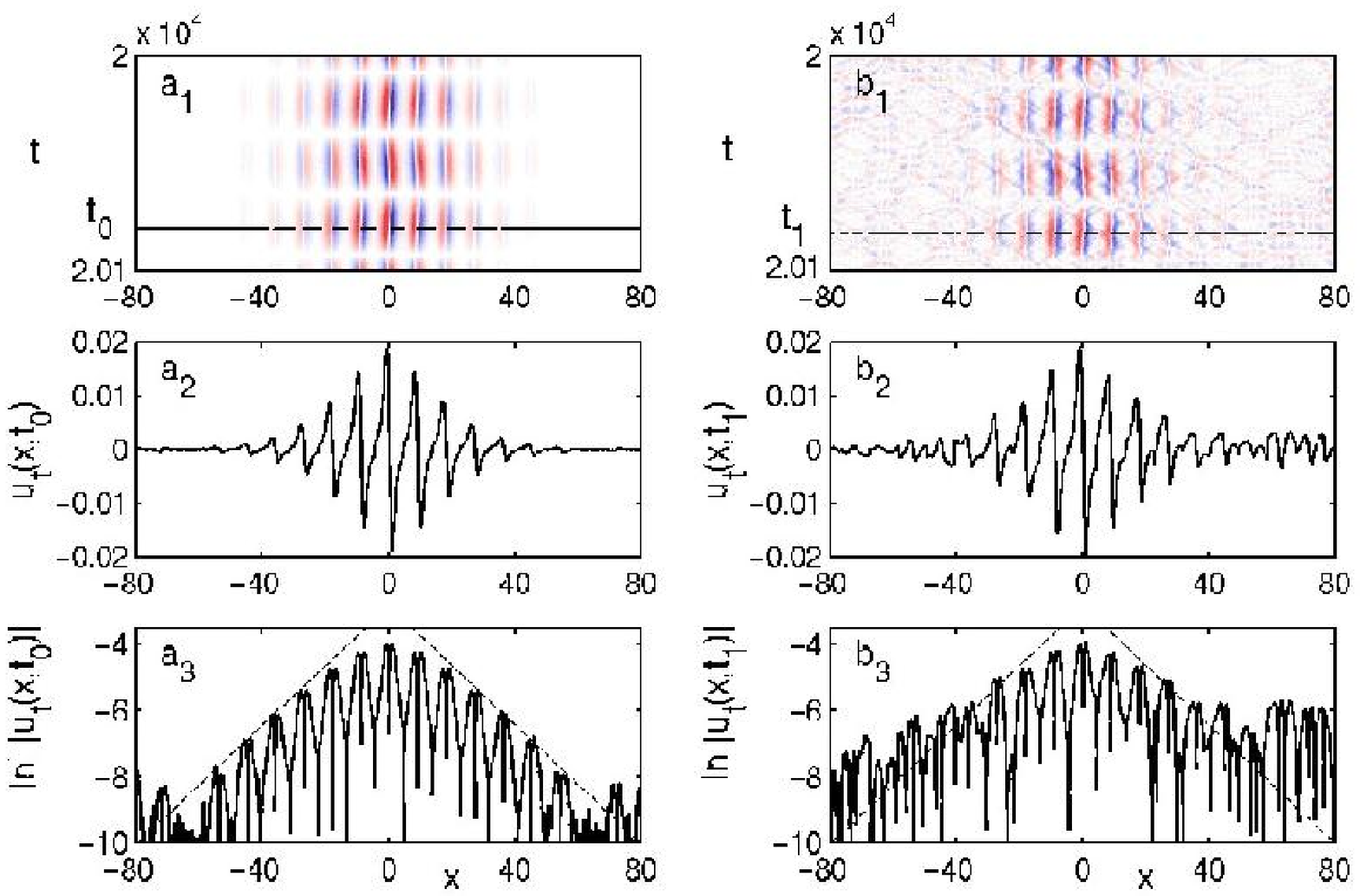}{\FigSize}{Robustness of the perturbed 
localized breather in the Gross-Pitaevskii equation: 
a) unperturbed and b) perturbed.
a$_1$ and b$_1$ depict a density plot of $|u_t(x,t)|$
---darker areas correspond to larger $u(x,t)$ variations. 
Insets a$_2$ and a$_3$ depict $u_t(x,t_0)$ and $\ln(|u_t(x,t_0)|)$
at $t=t_0$ for the unperturbed system, whereas
insets b$_2$ and b$_3$ depict the corresponding quantities
at $t_1$ for the perturbed systems.
%at $t=t_0$ for the unperturbed breather while b$_2$ and b$_3$ depict 
%the averaged solutions for the perturbed breather between $t_1$ and $t_2$.
%The times $t_0$, $t_1$ and $t_2$ are depicted in a$_1$ and b$_1$.
Same parameter values as in Fig.\ \ref{fig3.ps}.}
%%%%%%%%%%%%%%%%%%%%%%%%%%%%%%%%%%%%%%%%%%%%%%%%%%%%%%%%%%%%%%%%%%%%%%%%%

Localized breathers represent an important structure in the dynamics
of the GPE equation describing a mechanism whereby energy and information 
can be pinned down on the periodic potential lattice.
The occurrence of localized
breathers in weakly coupled oscillatory units is a common phenomenon.
Indeed, the existence of localized breathers
has been formally established in general
nonlinear Hamiltonian lattices of weakly interacting oscillators
\cite{MacKay:94-Flach:95}
by continuation from the so called anticontinuum
limit corresponding to the uncoupled case ($R=+\infty$) 
\cite{Aubry:94}.
%\cite{MacKay:94-Flach:95-Aubry:94}.
%
%A crucial item that deserves elaborating at this point is the
%structural stability of the homoclinic tangle 
%with respect to parameter changes and external perturbations. 
%
{\mybf 
The structural stability of the homoclinic tangle, arising from the
transverse nature of the intersections of the stable and unstable manifolds,
implies the persistence of the breather solution of the 2D map (\ref{2DM})
under parameter changes, in particular insuring the existence of
the breather solution for the GPE.
This persistence, together with the dynamic stability of the steady state
solution ($A_n=0$), ensures the existence of the breather solution
for the GPE.
}
Therefore, despite the various approximations used in the reduction of the
GPE to the 2D map (\ref{2DM}), we observe localized breather oscillations 
in the original GPE dynamics (Fig.\ \ref{fig3.ps}) for 
the predicted parameter values. 
More surprising is the robustness of the breathing dynamics
to significant perturbations under the GPE dynamics
(Fig.\ \ref{fig4.ps.jpg.ps}). 

To demonstrate this robustness we construct the initial configuration
(IC) as a concatenation of solitons with velocities, heights, widths
and positions predicted by our analysis, and with a $\pi$ phase shift 
between consecutive solitons (as in Fig.\ \ref{fig3.ps}).
We perturb each soliton in the chain by adding 
a random value to each of the initial velocities, heights, widths, positions 
and phases. The bounds for the perturbations correspond to 
\begin{itemize}
\itemsep 0pt
\item[a)] 15\% of the {\em maximal} velocity on velocities,
\item[b)] 10\% on heights,
\item[c)] 10\% on widths, 
\item[d)] 15\% of the {\em maximal} amplitude 
          ($0.15\!\times\! A_0$) on positions and
\item[e)] 15\% of $\pi$ on phases. 
\end{itemize}
Note that the perturbations on positions and velocities are proportional 
to the collective maximum and not to the individual values. 
After adding the perturbations we concatenate the solitons and
integrate the NLS (\ref{NLSE}) using a pseudo-spectral method. 
Additionally, we include  
\begin{itemize}
\itemsep 0pt
\item[f)] 5\% (stationary) perturbation to the potential and 
\item[g)] a noise level of $10^{-5}$ to $u(x,t)$ at {\em every} 
          time step of the integration.
\end{itemize}
%
%a noise level of 
%$10^{-5}$ to $u(x,t)$ at {\em every} time step of the integration
%(total run is $2.01\!\times\! 10^6$ iterations).
The total run is about 
$2\!\times\! 10^6$ iterations.
%$2.01\!\times\! 10^6$ iterations.
Fig.\ \ref{fig4.ps.jpg.ps} depicts the evolution of $u_t(x,t)$
for the unperturbed (left) and the perturbed (right) breather 
from direct simulations of the GPE equation.
Despite the large perturbation to the IC and the additive noise,
after a brief transient, the breather settles down and
retains its localization with an approximate exponential decay (inset b$_3$).
%The solutions depicted in insets b$_2$ and
%b$_3$ correspond to {\em averaged} configurations between $t_1$ and $t_2$
%in order to smooth the perturbations and noise.
%We also observe that, due to the perturbations, the synchronization 
%of the oscillations is lost outside the localized region.
It is interesting to note that stronger perturbations to $u$ 
did not necessarily destroy the localization but often resulted in a slow 
excursion of the localized region along the lattice.
The possibility of breather mobility in nonlinear lattices triggered 
by perturbations has been investigated previously \cite{Aubry:96}.

The robustness of the localized breather dynamics presented above
opens the possibility for experimental corroboration.
The large perturbations to the IC result in an initial transient involving 
radiative losses, which would correspond experimentally to a small number 
of atoms being spilled away from the central cloud and absorbed
by the non-condensed atoms at the periphery. After the transient,
the breather settles down and retains its localization with an
approximate exponential decay (inset b$_3$). While the breathers we
constructed are robust to a wide host of perturbations,
breathers are non-generic in the sense that arbitrary
initial conditions do not necessarily produce a breather.
Also, with large enough perturbations the breathing phenomena are
entirely destroyed.

%%%%%%%%%%%%%%%%%%%%%%%%%%%%%%%%%%%%%%%%%%%%%%%%%%%%%%%%%%%%%%%%%%%%%%%%%
\oneFIG{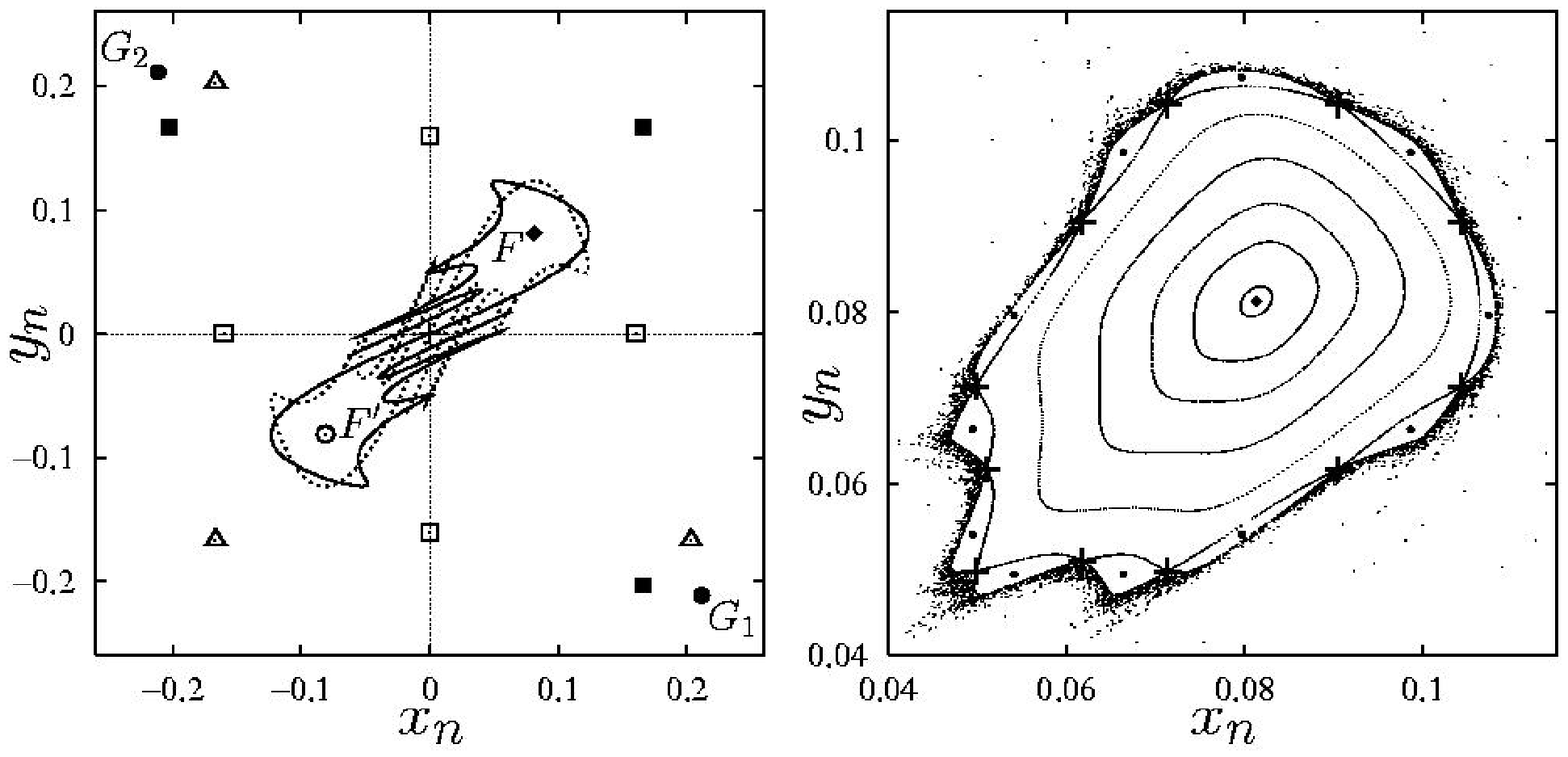}{\FigSize}{Typical phase space for
the reduced 2D map (\ref{2DM}). Left: some periodic orbits and
the homoclinic tangle. 
$F$ and $F'$ are fixed points, $\{G_1,G_2\}$ is
a period 2 orbit. Some higher order periodic 
orbits: period 3 (triangles and filled squares) and period 4 (open
squares). Right: behavior near the fixed point $F$ (diamond). The map
displays quasiperiodic orbits which disappear at the separatrix 
originating from the manifolds of a period 11 orbit (crosses). Outside 
this separatrix a single chaotic orbit is depicted (small dots).}
%%%%%%%%%%%%%%%%%%%%%%%%%%%%%%%%%%%%%%%%%%%%%%%%%%%%%%%%%%%%%%%%%%%%%%%%

%It is important to mention that, while
%an initial condition sufficiently close to a breather 
%does produce one after a transient evolution involving radiation loss, 
%localized breathers are non-generic in the sense that arbitrary initial 
%data for (\ref{NLSE}) does not necessarily produce a breather. 
%The radiation loss corresponds to a small number of atoms being
%spilled away from the central cloud and absorbed by the non-condensed
%atoms at the periphery.

We may use the dynamical reduction described above to devise other 
breathing phenomena of (\ref{NLSE}). Within our approximations, any 
bounded nontrivial orbit of the 2D map $M$ (\ref{2DM}) gives rise to 
complex oscillatory behavior of the condensates in (\ref{NLSE}). In 
particular, periodic points of $M$ correspond to {\em global} oscillations.  
The map $M$ has three fixed points: the origin, $F=(x^*,x^*)$ 
and $F'=(-x^*,-x^*)$ 
(see Fig.\ \ref{fig5.ps.jpg.ps}),
where $x^*=\sqrt{(2-a)/b}$. 
The fixed point at the origin gives 
rise to the trivial stationary solution $A_n$'s$\,=0$. 
The fixed points
$F$ and $F'$ yield solutions in which all condensates oscillate in phase 
with the same amplitude $x^*$. The corresponding global breather for the 
original system (\ref{NLSE}) is depicted in 
Fig.\ \ref{fig6a.ps}.a. 
Other interesting orbits arise from higher order periodic points. For 
example, the period two orbit $G_2=-M(G_1)=M^2(G_2)$ 
%$(G=(\hat x,-\hat x) {\rm\ with\ }\hat x^2 = {-(2+a)/b})$
(see Fig.\ \ref{fig5.ps.jpg.ps}) 
where $G_1=(\hat x,-\hat x)$ and $G_2=(-\hat x,\hat x)$ with
$\hat x = \sqrt{-(2+a)/b}$. This period two orbit 
corresponds to an amplitude configuration 
$\{\dots,\hat x,-\hat x,\hat x,-\hat x,\dots\}$,
ie.\ anti-phase oscillations 
(see Fig.\ \ref{fig6a.ps}.b).
It is in principle possible to construct more complex patterns for the
global oscillations from higher order periodic orbits of the 2D map
(cf.\ Fig.\ \ref{fig5.ps.jpg.ps}).

The 2D map also predicts global oscillations which are quasi-periodic 
in site index $n$. These orbits exist near the fixed point $F$
(see Fig.\ \ref{fig5.ps.jpg.ps}, right panel). 
The corresponding breather for the full periodic NLSE (\ref{NLSE}) has an 
in-phase global oscillation with a small modulation of the 
amplitudes
(given by the rotation number of the quasi periodic 
orbit around $F$).
An interesting possibility for the dynamics of coupled
condensates is the prospect of chaotic evolution. In a neighborhood of
the fixed point $F$ there is a region containing chaotic
orbits corresponding to chaotic oscillations for the condensates
(see right panel in Fig.\ \ref{fig5.ps.jpg.ps}). 
This corresponds to chaotic oscillations for the condensates 
(see Fig.\ \ref{fig7.ps}.c).
It should be possible, in principle, to find the onset of chaos
for the condensates by analyzing in more detail the reduced dynamics
on (\ref{2DM}).

%%%%%%%%%%%%%%%%%%%%%%%%%%%%%%%%%%%%%%%%%%%%%%%%%%%%%%%%%%%%%%%%%%%%%%%%%
\twoFIG{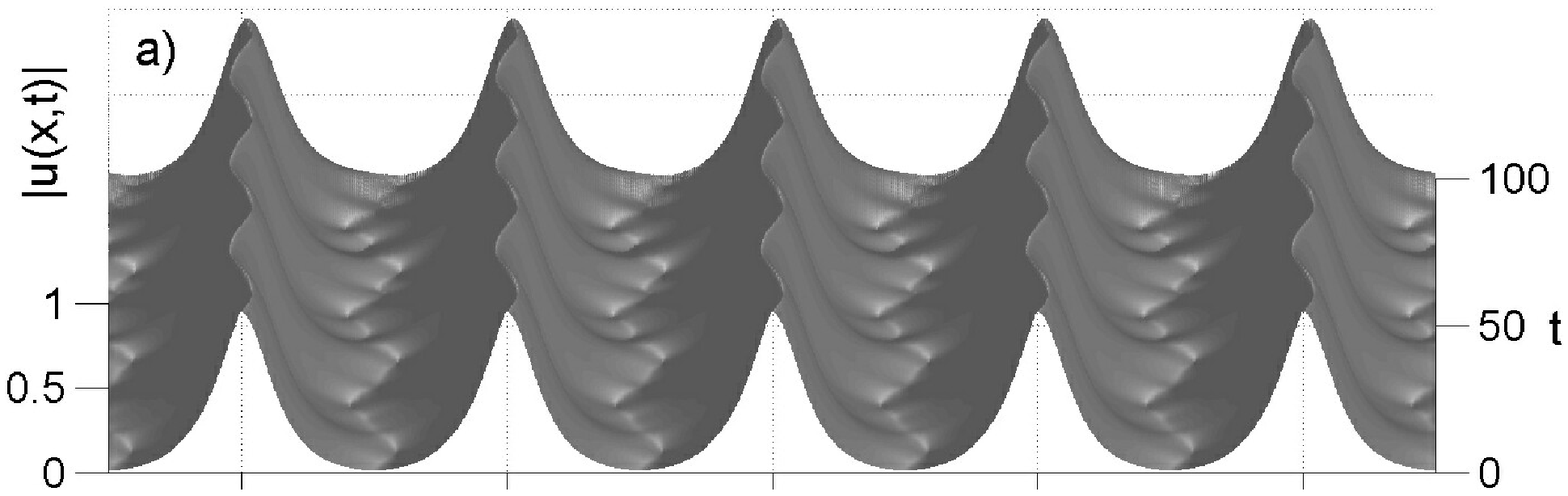}{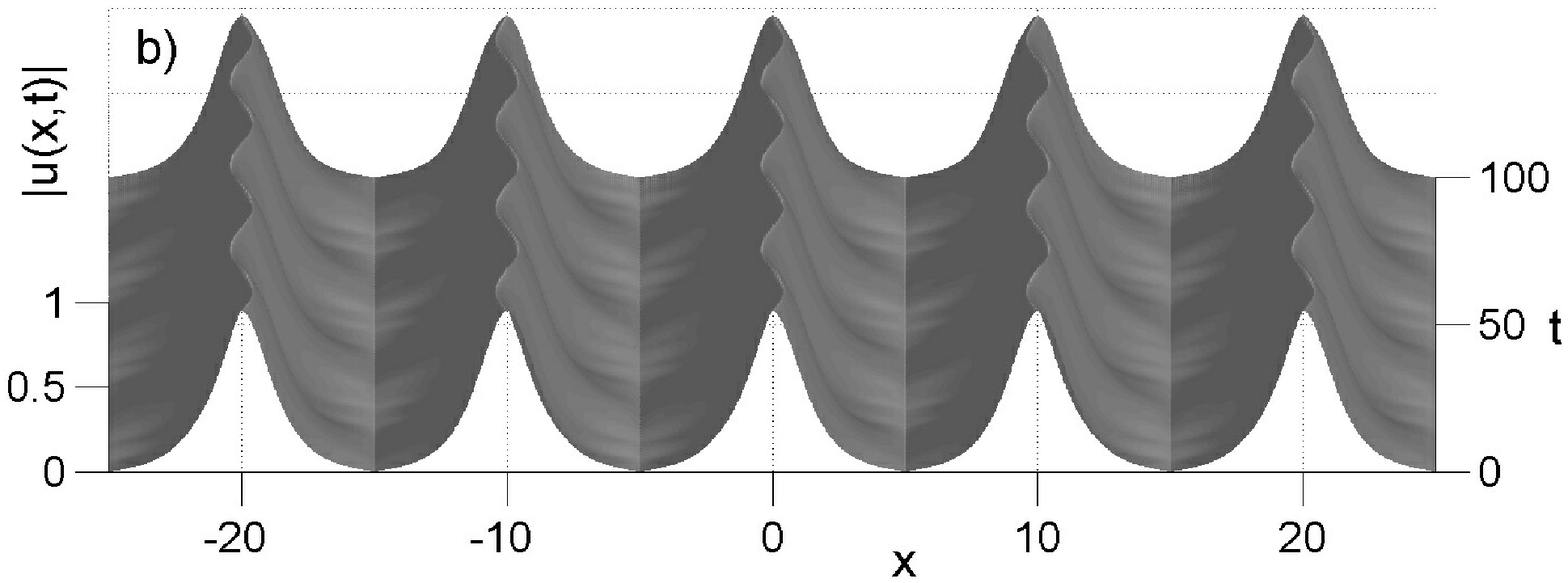}
{\FigSize}{0.2cm}{Global oscillations of the condensates
from simulations of the GPE equation. These global oscillations
arise from periodic points of the reduced dynamics (\ref{2DM}). 
a) In-phase oscillations corresponding to the
nontrivial fixed point $F$ (see Fig.\ \ref{fig5.ps.jpg.ps}). 
b) Anti-phase oscillations corresponding to a period two cycle of 
(\ref{2DM}) with alternating amplitude sign.}
%%%%%%%%%%%%%%%%%%%%%%%%%%%%%%%%%%%%%%%%%%%%%%%%%%%%%%%%%%%%%%%%%%%%%%%%%

%%%%%%%%%%%%%%%%%%%%%%%%%%%%%%%%%%%%%%%%%%%%%%%%%%%%%%%%%%%%%%%%%%%%%%%%%
\oneFIG{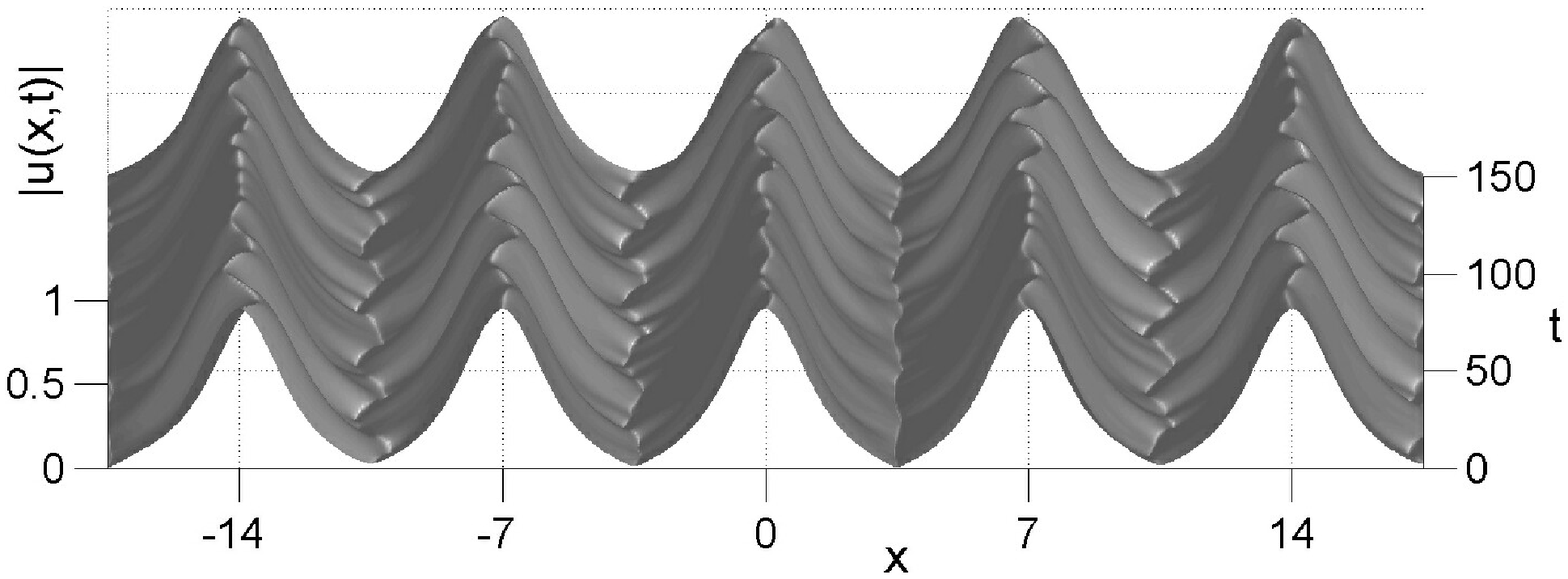}{\FigSize}{Chaotic global oscillations of 
the condensates from GPE dynamics corresponding to the chaotic orbit 
of the reduced map (\ref{2DM}) (see  Fig.\ \ref{fig5.ps.jpg.ps}).}
%%%%%%%%%%%%%%%%%%%%%%%%%%%%%%%%%%%%%%%%%%%%%%%%%%%%%%%%%%%%%%%%%%%%%%%%%

We have constructed a variety of novel global and localized oscillatory 
behaviors of BECs in periodic potentials, identifying these solutions 
with orbits of a reduced 2D map. 
A key ingredient of 
the construction of localized oscillations is the existence
of a homoclinic tangle. We demonstrate the surprising robustness of 
these solutions to perturbations. 
Since BEC experiments are quite delicate,
we do not expect that direct manipulation could produce an exact initial 
condition corresponding to a localized oscillation. Nonetheless, we believe
that localized oscillations may be observed in weakly coupled condensates
which are appropriately engineered and then permitted to radiate away
spurious energy.
The techniques presented here can, in principle, be extended to lattices 
in higher dimensions such as vortex lattices.
This has possible implications to modeling the interactions of 
atoms in optical traps that could potentially be used for quantum 
computing \cite{Jessen:00}.

We are grateful to J.N.\ Kutz, B.\ Deconinck, P.G.\ Kevrekidis,
P.\ Engels and W.P.\ Reinhardt
for providing stimulating discussions and relevant references. 
The authors acknowledge support from the Pacific Institute for the
Mathematical Sciences and NSERC grant \#611255 during the
completion of this work.

%\null\vskip -0.5cm 


\begin{thebibliography}{10}
\bibitem{RCG:note2}
Collapse for attractive BECs is identified with the
collapse of the NLS in dimensions $D\geq 2$. 
Since the NLS does not blow up in 1D, designing a trap that 
confines the BEC to an almost perfect 1D profile (a very thin cigar
shape) could arrest, or at least delay, collapse.
\end{thebibliography}
\end{document}